\documentclass[doublecol]{epl2}

\usepackage{amssymb}
\usepackage{amsmath}
\usepackage{graphicx}
\newcommand{\be}{\begin{equation}}
\newcommand{\ee}{\end{equation}}
\newcommand{\ben}{\begin{eqnarray}}
\newcommand{\een}{\end{eqnarray}}

\title{A note on the infrared behavior of the compactified Ginzburg--Landau model in a magnetic field}

\author{C.A. Linhares\inst{1}, A. P. C. Malbouisson\inst{2} 
 and M.L. Souza\inst{2}}

\institute{\inst{1}Instituto de F\'{\i}sica, Universidade do Estado do Rio de Janeiro,\\
Rua S\~{a}o Francisco Xavier, 524, 20559-900 Rio de Janeiro, RJ, Brazil\\
\inst{2}Centro Brasileiro de Pesquisas F{\'i}sicas, MCT,
22290-180, Rio de Janeiro, RJ, Brazil}

\pacs{11.10.Kk}{Field theories in dimensions other than four}
\pacs{11.30.Qc}{Phase transition}
\pacs{11.10.Wx}{Finite-temperature field theory}

\abstract{We consider the Euclidean large-$N$ Ginzburg--Landau model in $D$
dimensions, $d$ ($d\leq D$) of them being compactified.  
For $D=3$, the system can
be supposed to describe, in the cases of $d=1$, $d=2$, and $d=3$,
respectively, a superconducting material in the form of a film, of an
infinitely long wire having a rectangular cross-section and of a brick-shaped
grain. We investigate the fixed-point structure of
the model, in the  presence of an external magnetic field. 
An infrared-stable fixed points is found, which is independent of the number of compactified dimensions. This generalizes previous work for type-II superconducting films.}

\begin{document}

\maketitle

\section{Introduction}

A large amount of work has already been done on the Ginzburg--Landau (GL)
model, both in its single component and in the $N$-component versions, using
the renormalization group approach~\cite
{affleck,lawrie,lawrie1,brezin,radz,flavio,malbo}. In particular, an
analysis of the renormalization group in finite-size geometries can be found
in~\cite{cardy} and a general study of phase transitions in confined
systems is in~\cite{livro}. These studies have been performed to take into
account boundary effects on thermodynamical quantities, in particular on the transition temperature .
The existence of phase transitions are in this case associated to some
spatial parameters related to the breaking of translational invariance, for
instance, the distance $L$ between planes confining the system. Also, in
other contexts, the influence of boundaries on the behavior of systems
undergoing transitions have been investigated as in for instance~\cite{fadolfo}.

We  analyze in the present note effects of boundaries on the
transition by considering that such confined systems are modeled by
compactifying spatial dimensions~\cite{livro}. Compactification is 
engendered as a generalization of the Matsubara (imaginary-time)
prescription to account for constraints on the spatial coordinates. In the
original Matsubara formalism, time is rotated to the imaginary axis, $
t\rightarrow i\tau $, where $\tau $ (the Euclidean time) is limited to the
interval $0\leq \tau \leq \beta $, with $\beta =1/T$ standing for the
inverse temperature. The fields then fulfill periodic (bosons) or
antiperiodic (fermions) boundary conditions and are compactified on the $
\tau $-axis in an $S^1$ topology, the circumference of length $\beta $. Such
a formalism leads to the description of a system in thermal equilibrium at
the temperature $\beta ^{-1}$. Since in a Euclidean field theory space and
time are on the same footing, one can envisage a generalization of the
Matsubara approach to any set of spatial coordinates as well~\cite{polchinski,ademir}.

The  conceptual framework for studying simultaneously finite
temperature and spatial constraints has been developed by considering a
simply or nonsimply connected $D$-dimensional manifold with a topology of
the type $\Gamma _D^{d+1}={\rm{\bf R}}^{D-d-1}\times {\rm{\bf S} }^{1}_{0}\times {\rm{\bf S} 
}^{1}_{1}\times \cdots \times {\rm{\bf S} }^{1}_{d}$, with ${\rm{\bf S} }^{1}_{0}$
corresponding to the compactification of the imaginary time and $\,{\rm{\bf S} }
^{1}_{1},\dots ,{\rm{\bf S} }^{1}_{d}$ referring to the compactification of $d$
spatial dimensions~\cite{ademir}. Physical manifestations of this type of topology
include, for instance, the vacuum-energy fluctuations giving rise to the
Casimir effect (see for instance~\cite{livro} and other references therein). In the study of phase
transitions, the dependence of the critical temperature on the
compactification parameters is found in several situations of
condensed-matter physics~\cite{livro,amms,linhares,linhares1,jmp,jmp1}.
Also, this kind of formalism has been employed in the investigation of the
confining phase transition in effective theories for Quantum Chromodynamics~%
\cite{prd11,epl10,gnn,gnn2,npb09}. In the $\Gamma _D^{d+1}$topology, the
Feynman rules are modified by introducing a generalized Matsubara
prescription, performing the following multiple replacements
[compactification of a $(d+1)$-dimensional subspace]: 
\begin{eqnarray}
\int \frac{dk_0}{2\pi }\rightarrow \frac 1\beta \sum_{n_1=-\infty }^{+\infty
}\,,\;\;\;\;\int \frac{dk_i}{2\pi }\rightarrow \frac 1{L_i}\sum_{n_i=-\infty
}^{+\infty }\nonumber \\
k_0\rightarrow \frac{2(n_0+c)\pi }\beta
\;\;\;k_i\rightarrow \frac{2(n_i+c)\pi }{L_i}\;,  \label{Matsubara1}
\end{eqnarray}
where for each $i=1,2,\ldots ,d$, $L_i\,$is the size of the compactified
spatial dimension $i$ and $c=0$ or $c=1/2$ for, respectively, bosons and
fermions.

The compactification formalism described above has been applied to
field-theoretical models in arbitrary dimension with compactification of any subspace  ~\cite{jmp,jmp1,jmario}. This
formalism has also been developed from a path-integral approach in~\cite
{khanna1}. This allows to generalize to any subspace previous results in the
effective potential framework for finite temperature and spatial boundaries.
This mechanism generalizes and unifies results from recent work on the
behavior of field theories in the presence of spatial constraints~\cite
{fadolfo,ademir,jmario}, and previous results in the literature for
finite-temperature field theory as, for instance, in~\cite{gino}.

When studying the compactification of spatial coordinates, however, it is
argued in \cite{livro} from topological considerations, that we may have a
quite different interpretation of the generalized Matsubara prescription: it
provides a general and practical way to account for systems confined in
limited regions of space at finite temperature. Distinctly, we shall be
concerned here with a stationary field theory and employ the generalized
Matsubara prescription to study bounded systems by implementing the
compactification of spatial coordinates, no imaginary-time compactification
will be done. We will consider a topology of the type $\Gamma _D^d={\rm{\bf  R} 
}^{D-d}\times {\rm{\bf S} }^{1}_{1}\times {\rm{\bf S} }^{1}_{2}\times \cdots \times {\rm{\bf S} 
}^{1}_{d}$, where $\,{\rm{\bf S} }^{1}_{1},\dots ,{\rm{\bf S} }^{1}_{d}$ refer to the
compactification of $d$ {\em spatial} dimensions.

We consider in the present note the Euclidean vector $N$-component $
(\lambda \varphi ^4)_D$ theory at leading order in $1/N$, the system being
submitted to the constraint of being limited by $d$ pairs of parallel
planes. Each pair is orthogonal to the coordinate axes $x_1,\ldots ,x_d$,
respectively, and in each one of them the planes are at distances $
L_1,\ldots ,L_d$ apart from one another. 
This may be pictured as a parallelepiped-shaped box embedded in the $D$%
-dimensional space, whose parallel faces are separated by distances $L_1$, $%
L_2$,$\ldots $, $L_d$.
>From a physical point of view, we could  
take in particular $D=3$ and introduce temperature by means of the mass term
in the Hamiltonian in the usual Ginzburg--Landau way. These models can
then describe a superconducting material in the shapes of a film ($d=1$), of
a wire ($d=2$) and of a grain ($d=3$). With geometries such as these, some
of us have been able to obtain general formulas for the dependence of the
transition temperature and other quantities on the parameters delimiting the
spatial region within which the system is confined (see for instance~\cite
{jmp,jmp1} and other references therein).

We consider the critical behavior of the system under the influence of
an external magnetic field. Physically, for $D=3$, this corresponds to
superconducting films, wires and grains in a magnetic field. In~\cite{radz},
a large-$N$ theory of a second-order transition for arbitrary dimension $D$
is presented and the fixed-point effective free energy describing the
transition is found. The theory is based on the Ginzburg--Landau model with
the coupling of scalar and gauge fields. While ignoring gauge-field
fluctuations, the model includes an external magnetic field. The authors in~%
\cite{radz} also claim that it is possible that in the physical situation of 
$N=1$, a mechanism of reduction of the lower critical dimension could allow
a continuous transition in $D=3$. In \cite{malbo}, the possibility of the
existence of a phase transition for a superconducting film in the presence of
an external magnetic field has been investigated. This has been done in the
renormalization-group framework by looking for the existence of
infrared-stable fixed points for the $\beta $ function.

In this note, we study, for arbitrary space dimension $D$ and for any
number $d\leq D$ of compactified dimensions,
the fixed-point structure of the model, thus generalizing a previous 
 study for films~\cite{malbo}. We shall neglect the minimal coupling with the
vector potential corresponding to the intrinsic gauge fluctuations. 
Our main concern will be to analyze the model from
a field-theoretical point of view. In this sense, the present work may be seen 
as a further development of previous papers by some of us, 
as for instance~\cite{malbo,ademir,khanna1,linhares}.

\section{The compactified model in the presence of an external field}

We  consider the $N$-component vector model described by the
Ginzburg--Landau Hamiltonian density 
\begin{eqnarray}
{\cal H}&=&\left[ \left( \partial _\mu -ieA_\mu ^{\rm{ext}}\right) \varphi
_a\right] \left[ \left( \partial ^\mu -ieA^{\rm{ext,}\mu }\right) \varphi
_a\right] \nonumber \\
&&+m^2\varphi _a\varphi _a+u\,(\varphi _a\varphi _a)^2,
\label{hamiltoniana2}
\end{eqnarray}
in Euclidean $D$-dimensional space, where $u$ is the coupling constant and $
m^2$ is a mass parameter such that $m^2=\alpha \left( T-T_0\right) $ and $
T_0 $ the bulk transition temperature. Summation over repeated indices $\mu $
and $a$ is assumed. In the following, we will consider the model described
by the Hamiltonian (\ref{hamiltoniana2}) and take the large-$N$ limit, such
that $u\rightarrow 0$, $N\rightarrow \infty $ with $Nu=\lambda $ fixed. For $D=3$, 
from a physical point of view, such Hamiltonian
is supposed to describe type-II superconductors. In this case, it has been assumed 
that the external magnetic field ${\bf H}$ is parallel to the $z$-axis and
 the gauge ${\bf A}^{\rm{ext}}=(0,xH,0)$ was chosen. In the present $D$-dimensional 
case, we assume analogously a gauge ${\rm{A}}^{\rm{ext}}=(0, x_1H,0,0,\ldots ,0)$, with $\{x_i\}=x_1,x_2,\cdots x_D$, meaning that the applied external magnetic field lies on 
a fixed direction along one of the coordinate axis; for
simplicity, in the calculations that follow, we have adopted the notation $x_1\equiv x
$, $x_2\equiv y$. 

If we consider the system in unlimited space, the field $\varphi $ should be
written in terms of the well-known Landau-level basis, 
\begin{equation}
\varphi ({\bf r})=\sum_{\ell =0}^\infty \int \frac{dp_y}{2\pi }\int \frac{%
d^{D-2}p}{\left( 2\pi \right) ^{D-2}}\tilde{\varphi}_{\ell ,p_y,{\bf p}}\chi
_{\ell ,p_y,{\bf p}}({\bf r}),
\end{equation}
where $\chi _{\ell ,p_y,{\bf p}}({\bf r})$ are the Landau-level
eigenfunctions given by 
\begin{eqnarray}
\chi _{\ell ,p_y,{\bf p}}({\bf r})&=&\frac 1{\sqrt{2^\ell }\ell !}\left( \frac 
\omega \pi \right) ^{1/4}e^{i\left( {\bf p}\cdot {\bf r}+p_yy\right)
}e^{-\omega (x-p_y/\omega )^2/2}\nonumber \\
&&\times H_\ell \left( \sqrt{\omega }x-\frac{p_y}{%
\sqrt{\omega }}\right) ,
\end{eqnarray}
with energy eigenvalues $E_\ell \left( \left| {\bf p}\right| \right) =$ $%
\left| {\bf p}\right| ^2+\left( 2\ell +1\right) \omega +m^2$ and $\omega =eH$
is the so-called cyclotron frequency. In the above equation, ${\bf p} $ and $%
{\bf r}$ are ($D-2$)-dimensional vectors.
 We use Cartesian coordinates ${\bf r}=(x_1,\ldots
,x_d,{\bf z})$, where ${\bf z}$ is a $(D-d)$-dimensional vector, with
corresponding momenta ${\bf k}=(k_1,\ldots ,k_d,{\bf q})$, ${\bf q}$ being a 
$(D-d)$-dimensional vector in momentum space. Under these conditions, the generating
functional of correlation functions is written as
\begin{eqnarray}
{\cal Z}&=&\int {\cal D}\varphi ^{*}{\cal D}\varphi \,\exp \left(
-\int_0^{L_1}dx_1\cdots \int_0^{L_d}dx_d\right.\nonumber \\
&&\times \left.\int d^{D-d-2}z\;{\cal H}\left(|\varphi
|,|\nabla \varphi |\right) \right),
\end{eqnarray}
 the field $\varphi (x_1,\ldots ,x_d,{\bf z})$ satisfying the condition
of confinement inside the box, $\varphi (\{x_i\leq 0\},{\bf z})\;=\;\varphi
(\{x_i\geq L_i\},{\bf z})\;=\;$  const. Then the field representation
should be modified and have a mixed series-integral Fourier expansion of the
form 
\begin{eqnarray}
\varphi (x_1,\ldots ,x_d,{\bf z})=\sum_{\ell =0}^\infty
\sum_{i=1}^d\sum_{n_i=-\infty }^\infty c_{n_i}\int \frac{dp_y}{2\pi }\nonumber \\
\times\int
d^{D-d-2}{\bf q}\;b({\bf q})e^{-i\omega _{n_i}x\;-i{\bf q}\cdot {\bf z}}%
\tilde{\varphi}_\ell (\omega _{n_i},{\bf q}),  \label{Fourier2}
\end{eqnarray}
where, for $i=1,\ldots ,d$, $\omega _{n_i}=2\pi n_i/L_i$ and the
coefficients $c_{n_i}$ and $b({\bf q})$ correspond respectively to the
Fourier series representation over the $x_i$ and to the Fourier integral
representation over the ($D-d-2)$-dimensional ${\bf z}$-space. We now apply the 
Matsubara-like formalism according to Eq.~(\ref{Matsubara1}), remembering that here 
we have no imaginary-time compactification.

\section{Infrared behavior and fixed points}

In the following, we consider only the lowest Landau level $\ell =0$. For $%
D=3$, this assumption usually corresponds to the description of
superconductors in the extreme type-II limit. Under this assumption, 
we obtain  the effective $\left| \varphi \right| ^4$ interaction in
momentum space and at the critical point, $\lambda (p,D,\{L_i\};\omega )$,  from the four-point function, $\Gamma _D^{(4)}(p,\{L_i\},m=0)$. The four-point function is given by the sum of all chains of one-loop diagrams, which leads to
\begin{eqnarray}
\lambda (p,D,\{L_i\};\omega )\equiv \lim_{u\rightarrow 0\,;\;\,N\rightarrow \infty
}N\Gamma _D^{(4)}(p,\{L_i\},m=0)\nonumber \\
=\frac \lambda {1+\lambda \omega e^{-(1/2\omega
)(p_1^2+p_2^2)}\Pi (p,D,\left\{ L_i\right\} ,\,m=0 )},  \label{novoU}
\end{eqnarray}
with $Nu=\lambda $ fixed and where the single 1-loop bubble $\Pi (p,D,\{L_i\},m=0;\omega )$ is given by 
\begin{eqnarray}
\Pi (p,D,\{L_i\},m=0 ) =\nonumber \\
\frac 1{L_1\cdots L_d}\sum_{i=1}^d%
\sum_{n_i=-\infty }^\infty \int_0^1dx\int \frac{d^{D-d-2}q}{(2\pi )^{D-d-2}}
\nonumber \\
\times \frac 1{\left[ {\bf q}^2+\omega _{n_1}^2+\cdots +\omega
_{n_d}^2+p^2x(1-x)\right] ^2}.  \nonumber \\
  \label{novoPi}
\end{eqnarray}
This is the same kind of expression that is encountered in~\cite{jmp1}, with the only modification that $D\rightarrow D-2$ and the role of the mass is played by the quantity 
$\sqrt{p^2x(1-x)}$. Also, one should be reminded that $p$ is  a ($D-2$)-dimensional vector.  
The analysis is then performed along the same lines as in~\cite{jmp1} and we obtain,
analogously, 
\ben
\Pi(p,D,\{L_i\}, m=0) 
=\left( 2\pi \right) ^{1-D/2}\,\left[
2^{1-D/2}\frac 1{\left( 2\pi \right) ^2}c(D)\right. \nonumber \\
\left.\times \Gamma \left( 3-\frac D2\right)
\left( p^2\right) ^{D/2-3} \right. \nonumber \\
\left.+\int_0^1dx\,\left[ \sum_{i=1}^d\sum_{n_i=1}^\infty \left( \frac{\sqrt{%
p^2x(1-x)}}{2\pi L_in_i}\right) ^{(D-2)/2-2}\right. \right. \nonumber \\
\times\left. K_{(D-2)/2-2}\left( \frac 1{2\pi }\sqrt{
p^2x(1-x)}L_in_i\right) \right.  \nonumber \\
\left.+2\sum_{i<j=1}^d\sum_{n_i,n_j=1}^\infty \left( \frac{\sqrt{p^2x(1-x)%
}}{2\pi \sqrt{L_i^2n_i^2+L_j^2n_j^2}}\right) ^{(D-2)/2-2} \right. \nonumber\\
\left. \times K_{(D-2)/2-2}\left( \frac 1{%
2\pi }\sqrt{p^2x(1-x)}\sqrt{L_i^2n_i^2+L_j^2n_j^2}\right) +\cdots  
\right. \nonumber \\
\left. +2^{d-1}\sum_{n_1,\ldots ,n_d=1}^\infty \left( \frac{\sqrt{p^2x(1-x)%
}}{2\pi \sqrt{L_1^2n_1^2+\cdots +L_d^2n_d^2}}\right) ^{(D-2)/2-2} \right. \nonumber\\
\left.\times K_{(D-2)/2-2}\left( 
\frac 1{2\pi }\sqrt{p^2x(1-x)}\sqrt{L_1^2n_1^2+\cdots +L_d^2n_d^2}\right)
\right] , 
\nonumber \\
\label{novoPi-geral}
\een
where 
\begin{equation}
c(D) =\int_0^1dx\left( x(1-x)\right) ^{D/2-3}  
=2^{5-D}\sqrt{\pi }\frac{\Gamma\left(\frac{D}{2} -2\right)}{\Gamma\left(\frac{D-3}{2} \right)}.  
\label{def-b2}
\end{equation}

As for the infrared behavior of the $\beta $ function, it suffices to study
it in the neighborhood of $\left| p\right| =0$, so that we can  use the
asymptotic formula in the $|p|\approx 0$ limit,  for small values of the argument of the modified Bessel
functions, 
\begin{equation}
K_\nu (z)\approx \frac 12\Gamma (\nu )\left( \frac z2\right) ^{-\nu
}\;\;\;(z\sim 0).  \label{K}
\end{equation} 
>From Eq.~(\ref{novoPi-geral}), it turns out that in the $\left| p\right|
\approx 0$ limit, the bubble $\Pi $ is written in the form 
\begin{eqnarray}
\Pi (|p|\approx 0,D,\{L_i\},m=0 )\nonumber \\
=A(D)\left| p\right|
^{D-6}+C_d(D,\left\{ L_i\right\} ),
\label{Pi}
\end{eqnarray}
with 
\begin{equation}
A(D)=\left( 2\pi \right) ^{-D/2-1}2^{1-D/2}c(D)\Gamma \left( 3-\frac D2%
\right) ,  \label{A(D)2}
\end{equation}
and where the quantity $C_d(D,\left\{ L_i\right\} )$ is
\begin{eqnarray}
C_d(D,\left\{ L_i\right\} ) =\left( 2\pi \right) ^{-(D-2)/2}\nonumber \\
\times \int_0^1dx\,\left[ \sum_{i=1}^d\sum_{n_i=1}^\infty \left( \frac{\sqrt{%
p^2x(1-x)}}{2\pi L_in_i}\right) ^{(D-2)/2-2}\right.\nonumber \\
\times\left.K_{(D-2)/2-2}\left( \frac 1{2\pi }\sqrt{
p^2x(1-x)}L_in_i\right) \right.  \nonumber \\
+2\sum_{i<j=1}^d\sum_{n_i,n_j=1}^\infty \left( \frac{\sqrt{p^2x(1-x)%
}}{2\pi \sqrt{L_i^2n_i^2+L_j^2n_j^2}}\right) ^{(D-2)/2-2} \nonumber\\
\times K_{(D-2)/2-2}\left( \frac 1{%
2\pi }\sqrt{p^2x(1-x)}\sqrt{L_i^2n_i^2+L_j^2n_j^2}\right) +\cdots  
\nonumber \\
+2^{d-1}\sum_{n_1,\ldots ,n_d=1}^\infty \left( \frac{\sqrt{p^2x(1-x)%
}}{2\pi \sqrt{L_1^2n_1^2+\cdots +L_d^2n_d^2}}\right) ^{(D-2)/2-2} \nonumber\\
\left.\times K_{(D-2)/2-2}\left( 
\frac 1{2\pi }\sqrt{p^2x(1-x)}\sqrt{L_1^2n_1^2+\cdots +L_d^2n_d^2}\right)
\right].  
\nonumber \\ \label{Bd} 
\end{eqnarray}
If an infrared-stable fixed point exists for any of the models with $d$
confining dimensions, it is determined  by a study of the
infrared behavior of the Callan--Symanzik $\beta $ function, {\em i.e.}, in
the neighborhood of $|p|=0$. Therefore, we should investigate the above
equations for $|p|\approx 0$.

In this case, we consider a typical term in Eq.~(\ref{Bd}), which has the
form 
\begin{eqnarray}
\sum_{n_1,\ldots ,n_q=1}^\infty \left( \frac{\sqrt{p^2x(1-x)}}{2\pi \sqrt{%
L_1^2n_1^2+\cdots +L_q^2n_p^2}}\right) ^{(D-2)/2-s}\nonumber \\
\times K_{(D-2)/2-s}\left( \frac 1{2\pi }
\sqrt{p^2x(1-x)}\sqrt{L_1^2n_1^2+\cdots +L_q^2n_q^2}\right) ,
\label{typical}
\end{eqnarray}
with $s=2$ and $q=1,2,\ldots ,d$. 
In the limit $|p|\approx 0$, using Eq.(\ref{K}), Eq. (\ref{typical}) reduces to 
\begin{equation}
\frac 12\Gamma \left( \frac D2-s\right) E_q\left( \frac D2-s;L_1,\ldots
,L_q\right).  \label{G2}
\end{equation}
Eq.~(\ref{G2}) is expressed in terms of one of the multidimensional Epstein zeta
functions $E_q\left( \frac D2-s;L_1,\ldots ,L_q\right) $, for $q=1,2,\ldots
,d$, which are defined by~\cite{elizalde,elizalde1,elizalde2,kirsten} 
\begin{equation}
E_q\left( \nu ;\sigma _1,\ldots ,\sigma _q\right) =\sum_{n_1,\ldots
,n_q=1}^\infty \left[ \sigma _1^2n_1^2+\cdots +\sigma _q^2n_q^2\right]
^{\,-\nu }\;.  \label{EfuncS}
\end{equation}
Notice that, for $q=1$, $E_q$ reduces to the Riemann zeta function $\zeta
(z)=\sum_{n=1}^\infty n^{-z}$. 
One can also construct analytical continuations and recurrence relations for
the multidimensional Epstein functions, which permit to write them in terms
of modified Bessel and Riemann zeta functions \cite{ademir,elizalde1,elizalde2,kirsten}. One
gets 
\begin{eqnarray}
E_q\left( \nu ;L_1,\ldots ,L_q\right) =-\,\frac 1{2\,q}\sum_{i=1}^qE_{q-1}%
\left( \nu ;\ldots ,\widehat{L_i},\ldots \right) \nonumber\\
 +\,\frac{\sqrt{\pi }}{%
2\,d\,\Gamma (\nu )}\Gamma \left( \nu -\frac 12\right) \sum_{i=1}^q\frac 1{%
L_i}E_{q-1}\left( \nu -\frac 12;\ldots ,\widehat{L_i},\ldots \right) 
\nonumber \\
+\frac{2\sqrt{\pi }}{q\,\Gamma (\nu )}W_q\left( \nu -\frac 12,L_1,\ldots
,L_q\right) \;,  \nonumber \\
\label{Wd}
\end{eqnarray}
where the hat over the parameter $L_i$ in the functions $E_{q-1}$ means that
it is excluded from the set $\{L_1,\ldots ,L_q\}$ (the others being the $q-1$
parameters of $E_{q-1}$), and 
\begin{eqnarray}
W_q\left( \nu ;L_1,\ldots ,L_p\right) =
\sum_{i=1}^q\frac 1{L_i}%
\sum_{n_1,...,n_q=1}^\infty \nonumber \\
\left( \frac{\pi n_i}{L_i\sqrt{\cdots +\widehat{%
L_in_i^2}+\cdots }}\right) ^\nu \nonumber \\
\times K_\nu \left( \frac{2\pi n_i}{L_i}\sqrt{%
\cdots +\widehat{L_in_i^2}+\cdots }\right) \;,  \nonumber \\
\label{WWd}
\end{eqnarray}
with $\cdots +\widehat{L_in_i^2}+\cdots $ representing the sum $%
\sum_{j=1}^qL_j^2n_j^2\,-\,L_i^2n_i^2$.

Getting back to the infrared behavior, we  see from (\ref{G2}) that in the 
limit $|p|\approx 0$, the $p^2$-dependence of the modified Bessel functions exactly
compensates the one coming from the accompanying factors. Thus the remaining 
$p^2$-dependence is only that of the first term of (\ref{Pi}), which
is the same for all number of compactified dimensions $d$. 
It is worth mentioning also that  the simultaneous use of the Matsubara prescription and 
the Feynman parametrization leads to analiticity problems. This fact has been already investigated, 
for instance, in Refs~\cite{weldon,ayala}.
In our case, we see from Eqs.~(\ref{def-b2}) and (\ref{A(D)2}) that, due to the singularities of the gamma functions, $\Pi (|p|\approx 0,D,\{L_i\},m=0 )$ in Eq.~(\ref{Pi}) is well-behaved in the range of dimensions $4<D<6$. We will thus study our system for dimensions in this range. We emphasize that this range of dimensions, $4<D<6$, is the same that is compatible with the existence of a second-order phase transition for the system in  bulk form in previous publications~\cite{moore,radz,moore1}.

\subsection{Fixed points}

For all $d\leq D$, within the domain of validity of $D$, we have, by
inserting (\ref{Pi}) in Eq.(\ref{novoU}), the running coupling
constant 
\begin{eqnarray}
\lambda \left( |p|\approx 0,D,\{L_i\}\right) \approx  \nonumber \\
\frac \lambda {
1+\lambda \omega e^{-(1/2\omega
)(p_1^2+p_2^2)}\left[ A(D)|p|^{D-6}+C_d\left( D,\left\{ L_i\right\} \right)
\right] }.  \label{g3}
\end{eqnarray}
Let us take $|p|$ as a running scale, and define the dimensionless coupling 
\begin{equation}
g=\omega \lambda (p_1=p_2=0,D,\{L_i\})|p|^{D-6},  \label{g1(1)}
\end{equation}
where we remember that in this context $p$ is a $(D-2)$-dimensional vector.  

The $\beta $ function controls the rate of the
renormalization-group flow of the running coupling constant and  a
(nontrivial) fixed point of this flow is given by a (nontrivial) zero of the 
$\beta $ function. For $|p|\approx 0$, we obtain straightforwardly from
Eq. (\ref{g1(1)}),  
\begin{equation}
\beta (g)=|p|\frac{\partial g}{\partial |p|}\approx (D-6)\left[
g-A(D)g^{2}\right] ,  \label{beta2}
\end{equation}
>From Eqs.~(\ref{A(D)2}) and (\ref{def-b2}) we see that we have an infrared-stable fixed point, $g_{*}(D)$, for dimensions $D$ such that $4<D<6$, 
\begin{equation}
g_{*}(D)=\frac 1{A(D)}.  \label{gstar}
\end{equation}
We see that the $\{L_i\}$-dependent $C_d$-part of the subdiagram $\Pi $ does not
play any role in this expression and, since $A(D)$ is the same
for all number of compactified dimensions, so is $g_{*}$ only dependent on
the space dimension.

\section{Concluding remarks}

In this note, we have discussed the infrared behavior and the fixed-point structure of the 
$N$-component Ginzburg--Landau model in the large-$N$ limit, the system being
confined in a $d$-dimensional box with edges of length $L_i$, $i=1,2,\ldots
,d$ (compactification in a $d$-dimensional subspace).  We have studied the case in which  the system is submitted to the
action of an applied external magnetic field. In this case, we get the result that the existence of
an infrared-stable fixed point depends only on the space dimension $D$; it
does not depend on the number of compactified dimensions.

In the case of the system in the presence of an external magnetic field, it
is interesting to compare our results with those obtained for type-II
materials in bulk form. For instance, a large-$N$ analysis and a functional
renormalization-group study performed in Refs. \cite{moore,radz,moore1}
conclude for a second-order transition in dimensions $4<D<6$. The same
conclusion is obtained in Ref. \cite{flavio}. The authors of Ref.~\cite
{moore} claim, moreover, that the inclusion of fluctuations does not alter
significantly the main characteristic of the system, that is, the existence
of a continuous transition into a spatially homogeneous condensate. For the
system under the action of an external magnetic field, the existence of a
fixed point for $4<D<6$ should be taken as an indication, not as a
demonstration, of the existence of a continuous transition. As already
discussed in~\cite{moore,moore1}, in this case, even if infrared fixed
points exist, none of them can be completely attractive. The existence of an
infrared fixed point in the presence of a magnetic field, as found in this
paper, does not assure the (formal) existence of a second-order transition.
Anyway, we conclude that, for materials in the form of films, wires and
grains under the action of an external magnetic field, as is also the case
for materials in bulk form, if there exists a phase transition for $D<4$, in
particular in $D=3$, it should not be a second-order one.  Moreover, the 
fixed point is independent
of the size of the system or, in other words, the nature of the transition
in the presence of a magnetic field is insensitive to the confining geometry.

\section*{Acknowledgments}

A.P.C.M. acknowleges support from CNPq and FAPERJ (Brazil).

\end{document}